\documentclass[cites,graphicx]{article}
\usepackage{amsmath,cite,epsfig}

\title{Structure and mobility of cyclohexane as a solvent for
{\it Trans}$-$Polyisoprene}
\author{Roland Faller$^a$\\[3mm]
$^a$Department of Chemical Engineering,\\
University of Wisconsin, Madison, WI 53706-1691, USA \\
E-mail:faller@che.wisc.edu}
\begin{document}\maketitle
\renewcommand{\thefootnote}{\fnsymbol{footnote}}
%


\vspace{1ex}

\noindent Solutions of {\it trans}$-$polyisoprene in cyclohexane are
investigated in atomistic detail at different compositions at two different
temperatures. We investigate the influence of polymer concentration on the
dynamics of the solvent molecules and the structure of the solvation shell. The
double bonds along the polymer backbone are preferentially approached by the 
solvent molecules. The mobility of cyclohexane molecules decreases with increasing
polymer concentration at ambient conditions. The reorientation of molecules 
becomes more anisotropic with concentration as the polymer hinders the 
reorientation of the molecular plane. At elevated temperatures the influence of 
the polymer is weaker and the reorientation of the solvent is more isotropic. 
Additionally, a fast and efficient way to set up atomistic simulations is shown 
in detail in which the initial simulations increase in length and in the 
simulation time-step. The excess energy from initial overlaps is removed by 
resetting the velocities at regular intervals.

\vspace{1cm}
\section{Introduction}
Polymeric solutions are important for the understanding of technologically
important applications of polymers. Therefore they have been in the focus of
molecular simulations over the recent decade. Many studies have been performed
to investigate generic characteristics of polymer solutions by means of simple
bead-spring models~\cite{duenweg91,ahlrichs99}. However, to investigate the
behavior of a system on the local scale, fully atomistic simulations will be
required. Several authors have dealt with atomistically detailed descriptions
of polymer solutions~\cite{moe95,perico98,witt00,borodin01}. But, there are only
few studies how the polymer influences the solvent
dynamics~\cite{mplathe96,mplathe98b}. Especially, the region of low weight
fraction of polymer has not yet been studied. This contribution focuses on the
liquid structure and dynamics of small solvent molecules in the neighborhood
of oligomer chains in dilute solution. We use as an example
{\it trans}-polyisoprene dissolved in cyclohexane. Cyclohexane is a rather
spherical organic molecule and experimentally known to be a good solvent for
polyisoprene~\cite{tsunashima88,einaga94}. Polyisoprene by itself is one of the
most commonly used polymers.  Its natural form, cautschuk, is highly abundant in 
nature and for technological purposes, synthetic variations are easily 
industrially polymerised from isoprene. Still, the solutions of such important
materials are not yet completely understood from a microscopic viewpoint. 
\section{Details of the System}
Systems containing one or two 15-mers (C$_{75}$H$_{122}$) of
{\it trans}-polyisoprene (PI) are simulated in a solution consisting of~250
or~500 molecules of cyclohexane (C$_6$H$_{12}$). An overview over all the
systems is given in Table~\ref{tab:systems}, also the resulting densities in 
the simulations are shown in that figure. Details of the self-developed
forcefields are found in ref.~\cite{faller01a} for {\it trans}-polyisoprene
and refs.~\cite{faller99c,schmitz99a} for cyclohexane. The polyisoprene
forcefield is in contrast to the original version enhanced with torsions for
the methyl groups with a strength of 3.3~kJ/mol per atom (9.9~kJ/mol per
methyl group). The cyclohexane forcefield is augmented with torsion potentials
for all the C$-$C bonds. Each of them has a strength of 10~kJ/mol in order to
prevent random flipping between configurations. The cyclohexane molecule can
occur in the so-called {\it chair} or {\it boat} configuration. They are
distinguished if the connection between C$_1$ and C$_4$ intersects the plane
defined by the other carbons ({\it chair}) or not ({\it boat}). All simulations
and analyses are performed using the YASP molecular dynamics
package~\cite{yasp}. No charges are used throughout the simulations.

\begin{table}
  \begin{center}
    \caption{\label{tab:systems} Details of the different systems. $N_P$ is the
     number of oligomers (C$_{75}$) and $N_C$ the number of cyclohexane
     molecules. $c$ is the concentration in weight \% polymer.
     $t_{\text{sim}}$ is the simulated time for the systems. Additionally the
     resulting densities $\rho$ are shown for the simulated temperatures.}
    \vspace{1ex}
    \begin{tabular}{ccccccc}
      \hline
      \# & $N_P$ & $N_C$ & c & $t_{\text{sim}}$ & $\rho(\text{300K})$ 
	& $\rho(\text{413K})$\\
      \hline
      0 & 0 & 250 & 0.0\% & 1~ns & 756.3 kg/m$^3$ & 627.4 kg/m$^3$  \\
      1 & 1 & 250 & 4.6\% & 2~ns & 764.2 kg/m$^3$ & 637.2 kg/m$^3$ \\
      2 & 1 & 500 & 2.4\% & 2~ns & 757.5 kg/m$^3$ & 629.2 kg/m$^3$ \\
      3 & 2 & 500 & 4.6\% & 2~ns & 762.5 kg/m$^3$ & 635.2 kg/m$^3$ \\
      4 & 2 & 250 & 8.9\% & 2~ns & 768.2 kg/m$^3$ & 643.5 kg/m$^3$ \\
      \hline			  
    \end{tabular}		  
  \end{center}			  
\end{table}

Both forcefields have been proven useful in recent simulations. The
polyisoprene forcefield was used to investigate the dynamics of oligomer
melts~\cite{faller01a} whereas the cyclohexane model was used in
simulations of pure cyclohexane and its mixtures with cyclohexene at ambient
conditions~\cite{schmitz99a}.
\section{How to set up and equilibrate an atomistic simulation}
In atomistic simulations, the way a system is initially set up and equilibrated
is crucial. Thus, we describe the equilibration procedure for our systems in
detail. For system~1 (one polymer chain solvated by 250 cyclohexanes) the 250
cyclohexane molecules were put onto a regular fcc lattice in a cubic box of
4~nm side-length corresponding to a density of $\rho=545.8$kg/m$^3$. An $NVE$
simulation with a very short timestep of $10^{-3}$~fs was started. The timestep
was increased from $10^{-3}$~fs to 1~fs in three steps; in each step it was
multiplied by a factor of 10. Each simulation was run for 1000 steps at
constant volume and energy. The velocities of the particles were reset from a
Maxwell-Boltzmann distribution corresponding to 300~K before each run. The
polyisoprene molecule was inserted at random, and the pre-equilibration
procedure restarted. This time the initial timestep was set to $10^{-6}$~fs and
the steps were increased by factors of 100, only the last two increases were by
factors of 10. The initial polyisoprene conformation was taken from the
C$_{75}$ chains of the melt system~3 of ref.~\cite{faller01a}. In the rare case
that one of the short simulations failed because of a non-convergence of the
SHAKE algorithm~\cite{ryckaert77} an additional reset of the velocities was
applied.

This method of pre-equilibration is very fast for a fully atomistic system. It
avoids the use of unphysical soft-core simulations. By resetting the velocities
to the desired temperature periodically, the system moves very efficiently away
from its initial high energy. One can randomly insert the solute if the density
is low and the timestep sufficiently short. The very short simulations at short
timesteps have the additional advantage that not too much energy is transferred
to internal degrees of freedom which would lead to longer simulations in
later equilibration stages.

After the simulation was running stable under $NVE$ conditions at a timestep
of 1~fs, the Weak-Coupling themostat~\cite{berendsen84} was switched on. 10~ps
later  the isotropic Weak-Coupling constant pressure routine was switched on
to compress the system to the correct density. The compressibility
was set to the experimental value of 1.12~GPa$^{-1}$ of pure
cyclohexane~\cite{CRC}. Coupling times for pressure and temperature were
$\tau_T=0.2$~ps and $\tau_p=2.0$~ps respectively. The first constant pressure
run, however, started with $\Delta t = 0.1$~fs and $\tau_p=200$~ps. A stepwise
increase of the timestep and decrease of the coupling times was performed until
the simulations under production conditions were possible ($\tau_T=0.2$~ps,
$\tau_p=2.0$~ps, $\Delta t=2$fs).

A second system was produced the same way but independently with another 15-mer
from the melt simulations. These two systems were merged to form the systems
with two polymers. The system with only 250~cyclohexane molecules and
2~polymers was produced by taking out every second cyclohexane and letting the
system relax to the correct density in an $NpT$ simulation.

After proper relaxation at the correct timestep using temperature and pressure
coupling, the production runs were started. The relaxation lasted longer than
the correlation time for the end-to-end length of the polymer to lose the
memory of the melt structure. The production runs lasted up to 5~ns. The
reorientation correlation time of the chains was around 2~ns. Configurations
were saved every 1000 steps (i.e. 2~ps). The simulations were performed at
ambient conditions ($T=300\text{K}$, $p=101.3\text{kPa}$) and at
$T=413\text{K}$, $p=101.3\text{kPa}$. The higher temperature was chosen for 
consistency with simulations of solutions of polyisoprene in toluene~\cite{moe95}
although cyclohexane and toluene are gases at that temperature in nature. This 
vaporization transition is not reproduced by the force-field. The cutoff for the 
Lennard-Jones
interaction was set to $0.9\text{nm}$ and for the neighbor list~\cite{allen87}
to $1.0\text{nm}$. The neighbor-list was updated every 10 timesteps and the
total momentum of the box was reset to zero every 1000 timesteps. A pure
cyclohexane system was simulated to investigate the influences of the newly
added torsions in the forcefield in comparison to ref.~\cite{schmitz99a}.
\section{Statics and Dynamics of Cyclohexane solvating Polyisoprene}
\begin{figure}
\includegraphics[width=0.7\linewidth]{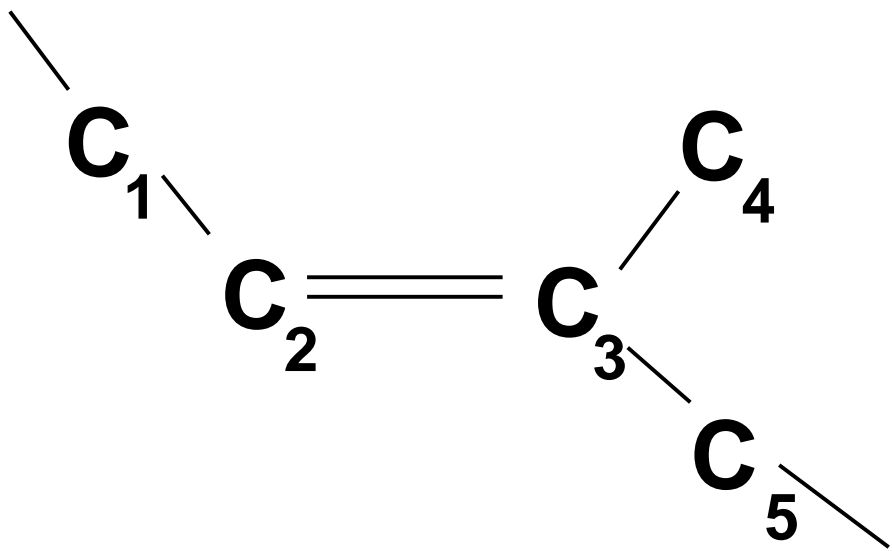}
\caption{Sketch of the carbons of a {\it trans}-polyisoprene monomer
indicating the numbers they are referred to  in this work}
\label{fig:sketch}
\end{figure}

In this section the carbons along the polymer chain are referred to in the
following way: C$_1$ is the non-methyl end of the monomer. C$_2$ and
C$_3$ are connected by the double bond. C$_2$ is connected to C$_1$, C$_4$ is
the methyl carbon, and C$_5$ is between C$_3$ and C$_1$ (cf. 
Fig.~\ref{fig:sketch}).

\begin{figure}
  \includegraphics[width=0.45\linewidth]{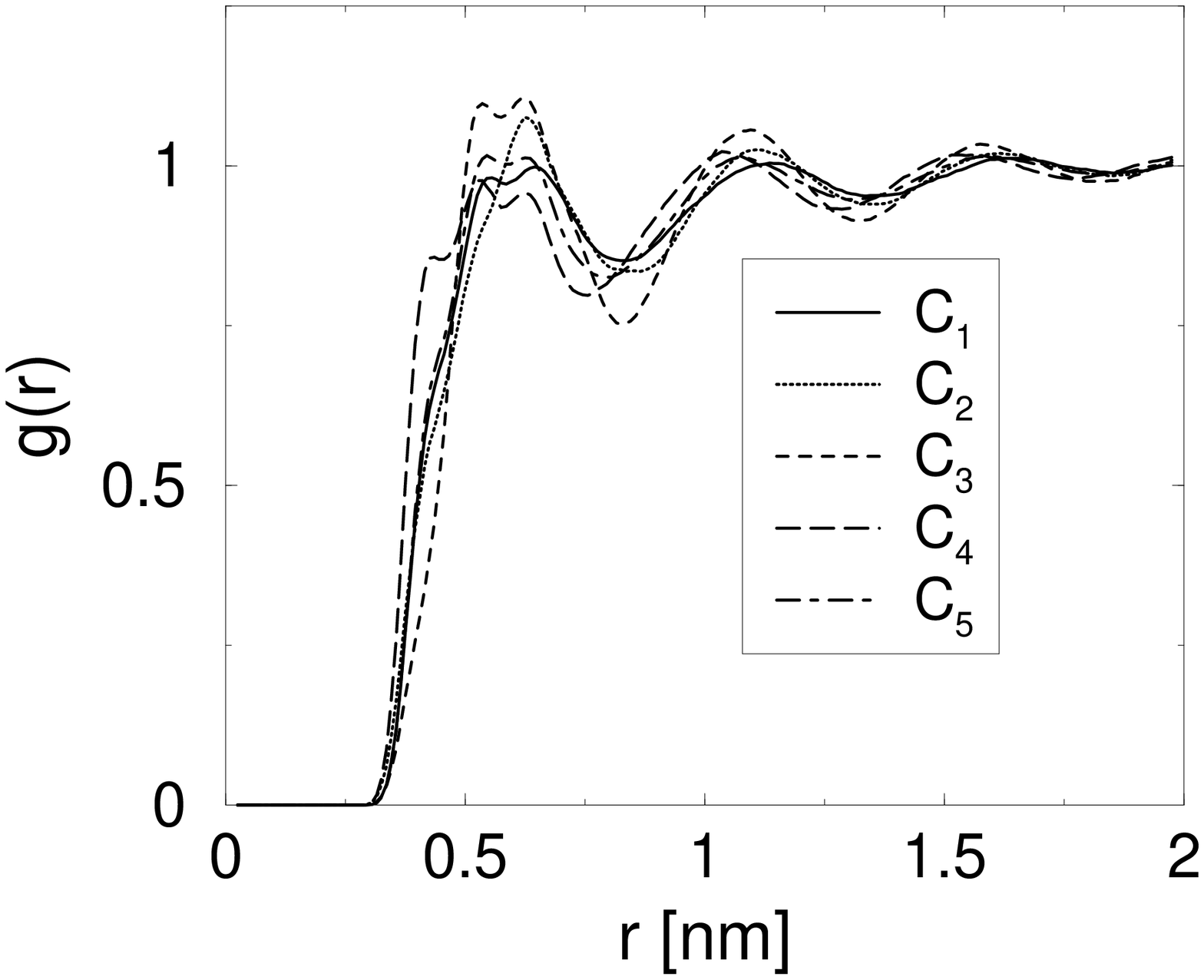}
  \includegraphics[width=0.45\linewidth]{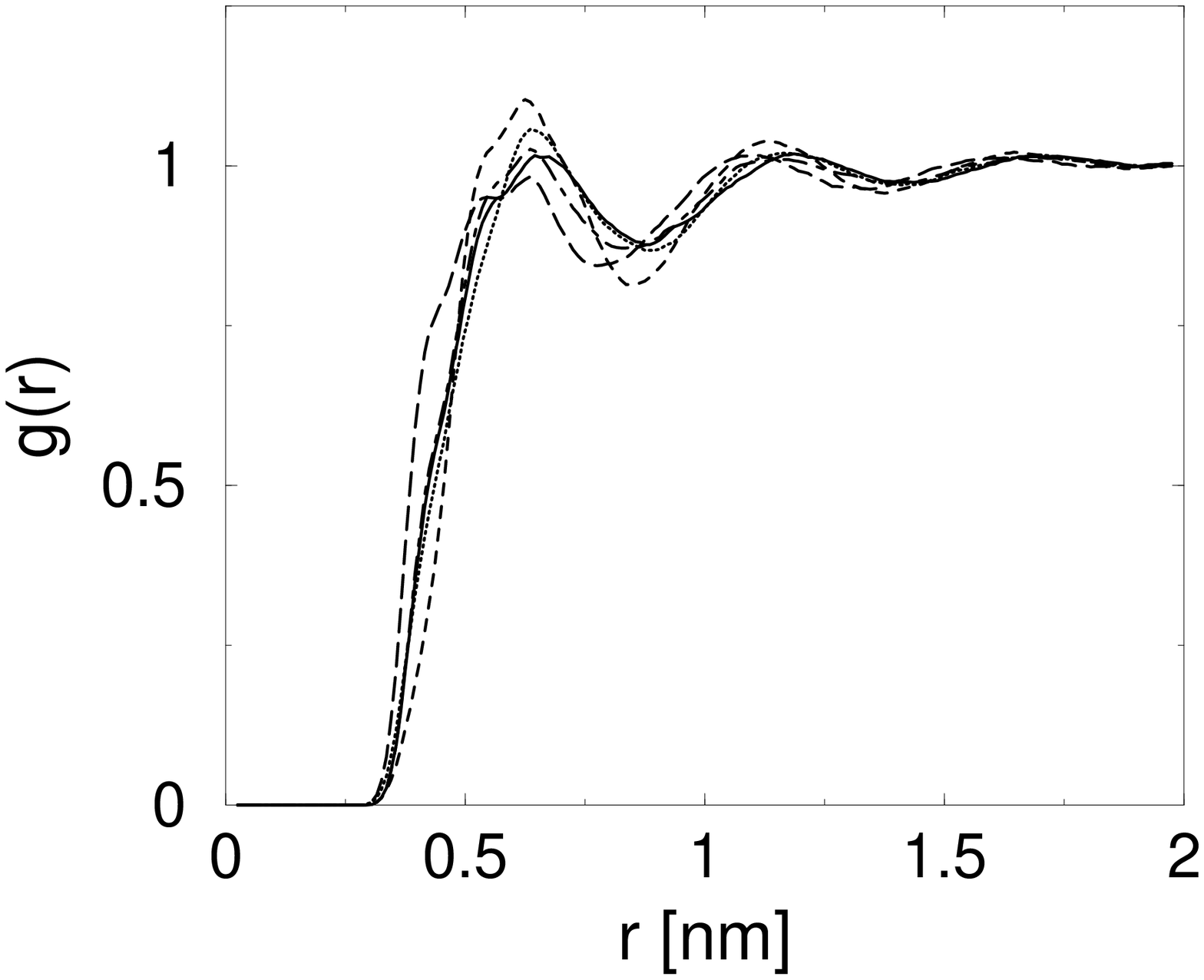}
  \caption{Radial distribution functions of the polymer carbons against the
  cyclohexane carbons for the system with 2 polymers and 500 cyclohexanes.
  a) 300K b) 413K. The legend of figure a) applies to both parts.}
  \label{fig:rdf}
\end{figure}

The first question we want to address is the way the cyclohexane molecules
solvate polyisoprene. Therefore we investigate the number of cyclohexane
carbons close to the five distinct carbons along the chain. We do not
distinguish between different concentrations here. To avoid end effects, the
end monomers are excluded from the calculation. The resulting radial
distribution functions are shown in Figure~\ref{fig:rdf}.

The main difference between the two temperatures is that for $\text{T}=300$~K
the neighbor peak is a double peak for some of the carbons. The two sub-peaks
are about 0.07~nm apart. This structure does not persist at the elevated
temperature. The overall height of the peaks is, however, unchanged. This
indicates that the number of cyclohexanes solvating a monomer is not
temperature dependent. The arrangement of the solvent molecules, however, is.
C$_4$ which is the side group, can come into closest contact with the solvent
visible at the shoulder at $r\approx0.4$~nm. It was also seen in the melt
investigations that this carbon is most exposed~\cite{faller01a}. All carbons
exhibit a peak at the first solvation shell at $r\approx0.55-0.6$~nm. A second
($r\approx1.1$~nm) and third ($r\approx1.55$~nm) solvation shell are visible.
The first peak is strongest for the two double bonded carbons. They are not
very strongly shielded by hydrogens and are therefore rather easily accessible.
This is in contrast to the melt simulations where the lacking mobility of
these carbons prevented a close approach to neighboring chains~\cite{faller01a}.
As the mobility of the solvent is much higher than that of the polymer, the 
differences in mobility of the polymer atoms is not important. At the higher 
temperature, the second and third solvation shell are slightly shifted to larger 
distances. The third solvation shell is rather broad.

The cyclohexane diffusion is found to be completely isotropic and the molecules
have diffused several times their gyration radius. The diffusion coefficient
is measured by fitting the mean-squared displacement of the center-of-mass of
the cyclohexane molecules to $\langle\vec{r}\,^2\rangle=6Dt$. The results are
found in Table~\ref{tab:diffu}. The diffusion coefficient for pure cyclohexane
in this model is found to be $0.78\times 10^{-5\,}\text{cm}^2/\text{s}$ at
$T = 300\text{K}$. The value of $0.89\times 10^{-5}\,\text{cm}^2/\text{s}$
reported in reference~\cite{schmitz99a} is about $15\%$ higher. This is probably
due to the missing internal torsions in that work which make it easier for the
molecule to squeeze through small voids. The diffusivity of cyclohexane
decreases with increasing polymer concentration. The polymer molecules can be
viewed as obstacles for the solvent. This decrease is opposite to the pronounced
increase of cyclohexane mobility in the presence of the smaller
cyclohexene~\cite{schmitz99a}. We can thus conclude that the mobility of
cyclohexane depends strongly on the size and concentration of the molecules with
which it is mixed.

\begin{table}
  \begin{center}
  \caption{ \label{tab:diffu}Motion of cyclohexane in the presence of
  polyisoprene. The error of the diffusion coefficients was estimated by the
  anisotropy of the mean-squared displacements. The errors of the
  reorientation times were estimated by the scatter of the times of the
  different vectors. The upper half of the table corresponds to T=300~K, the
  lower half to T=413~K. The $x-y$ notation means that the system contains $x$
  polymer chains and $y$ solvent molecules. The differences between the $1-250$ 
  and the $2-500$ system can be used as an estimate of the simulation error or
  finite size effects as both these systems are at the same concentration. These
  effects are on the order of less than 5\%.}

  \vspace{1ex}
  \begin{tabular}{ccccc}
  \hline
  conc. & D [$10^{-5}\text{cm}^2/\text{s}$]& $\tau_{\parallel}$ [ps]
    & $\tau_{\perp}$ [ps] & $\tau_{\perp}/\tau_{\parallel}$\\
  \hline
  pure  & 0.78 $\pm$ 0.09 & 3.2  $\pm$ 0.4  & 4.4 $\pm$ 0.1   & 1.38\\
  1-500 & 0.76 $\pm$ 0.02 & 3.84 $\pm$ 0.06 & 4.69 $\pm$ 0.05 & 1.22\\
  1-250 & 0.66 $\pm$ 0.03 & 3.65 $\pm$ 0.08 & 4.90 $\pm$ 0.06 & 1.34\\
  2-500 & 0.71 $\pm$ 0.03 & 3.83 $\pm$ 0.12 & 5.08 $\pm$ 0.14 & 1.32\\
  2-250 & 0.65 $\pm$ 0.03 & 3.84 $\pm$ 0.04 & 5.24 $\pm$ 0.06 & 1.36\\
  \hline
  \hline
  pure  & 5.2 $\pm$0.2 & 1.4  $\pm$ 0.1  & 1.45 $\pm$ 0.04 & 1.03\\
  1-500 & 5.6 $\pm$0.5 & 1.30 $\pm$ 0.02 & 1.43 $\pm$ 0.02 & 1.1\\
  1-250 & 5.0 $\pm$0.4 & 1.38 $\pm$ 0.02 & 1.41 $\pm$ 0.02 & 1.02\\
  2-500 & 5.3 $\pm$0.3 & 1.32 $\pm$ 0.02 & 1.42 $\pm$ 0.02 & 1.08\\
  2-250 & 4.7 $\pm$0.3 & 1.35 $\pm$ 0.04 & 1.46 $\pm$ 0.02 & 1.08\\
  \hline
  \end{tabular}
  \end{center}
\end{table}

Pure cyclohexane reorients faster in the plane defined by the carbon
ring than the plane flips. To investigate the reorientation behavior in
the presence of the polymer, we measure the two principal reorientation times
of an oblate molecule, i.e. $\tau_{\parallel}$ measures the reorientation of
vectors connecting carbons across the ring, whereas $\tau_{\perp}$ is the
characteristic time for rotating the axis normal to the plane. This axis is
defined in the simulations as cross-product of two C$-$C vectors across the
ring. The former measures the reorientation of the molecule in its plane,
whereas the latter measures the reorientation of the plane. In an earlier
study, we found that pure cyclohexane, as well as cyclohexane in a mixture with
cyclohexene, reorients faster in-plane than the plane
reorients~\cite{schmitz99a}. This is expected for a dense liquid; in the 
gas-phase, one expects this order to be reversed due to the moments of inertia. 
The reorientation times presented in Table~\ref{tab:diffu} were derived by an 
exponential fit to the first 25~ps of the correlation function at 
$\text{T}=300\text{K}$ (10~ps for $\text{T}=413$~K). These fits reproduced the 
curves very accurately. The reorientation times are measured as correlation times
of the first Legendre polynomial $P_1(t)=\langle\vec{u}(t)\vec{u}(0)\rangle$. The
resolution and accuracy is better than for $P_2(t)$ as the corresponding times 
are longer. $\vec{u}(t)$ is the unit vector we are monitoring. We find signs for 
Debye rotation $T_1=3T_2$, however, as $T_2$ is very short this cannot be 
completely decided. The cyclohexane model used in this work is slightly more 
anisotropic than its predecessor without internal torsions. The in-plane 
reorientation is exactly the same as in the old model~\cite{schmitz99a}. The 
reorientation of the plane is about 10\% slower as in the model lacking internal 
torsions.

The in-plane reorientation time is practically unaffected by the concentration
of the polymer except for pure cyclohexane, whereas the reorientation of the
plane becomes slower with increasing concentration. As the in-plane
reorientation does not need any other molecule to make room for the transition
to occur, the opposite is the case for the plane-flip. Therefore the polymers
act as obstacles to the molecules in their neighborhood and make their dynamics
more anisotropic as concentration increases. Pure cyclohexane appears to be a
different case as it is rather strongly anisotropic; this manifests itself in a
clearly faster in-plane reorientation than all the other cases. This suggests
that the in-plane reorientation of the molecules close to the polymer is hindered
by the chain. This effect is similar in strength for all concentrations.

\begin{figure}
  \begin{center}
    \includegraphics[angle=-90,width=0.7\linewidth]{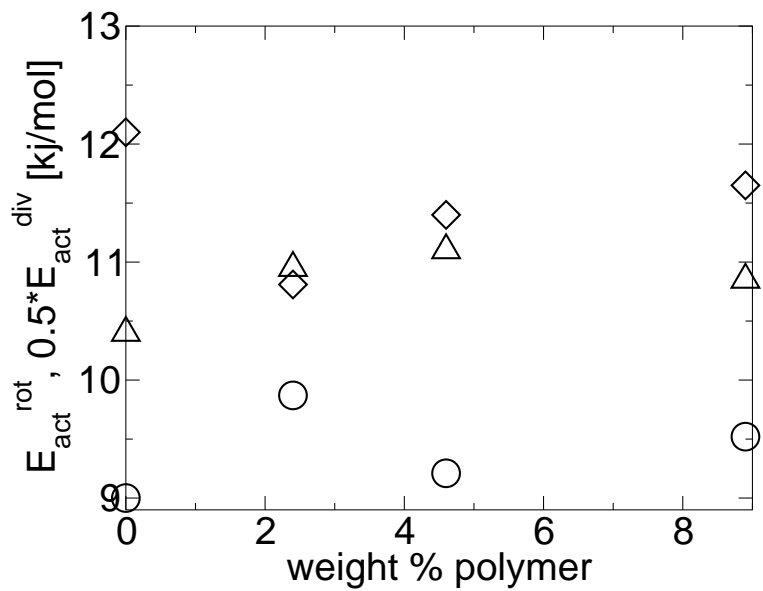}
    \caption{Estimated activation energies for the rotation in the plane
    (diamonds) and the plane flip (circles) depending on polymer concentration.
    Additionally the activation energy for the diffusion (triangles) is shown. 
    The values for diffusion are divided by a factor of 2.}
    \label{fig:arrh}
  \end{center}
\end{figure}

At the elevated temperature, the reorientation times are clearly shorter. The
anisotropy at this temperature becomes weaker, and in the range of
concentrations here, it does not depend systematically on polymer content.
Assuming an Arrhenius behavior, we can estimate activation energies for the
different rotations and the diffusion; they are shown in Figure~\ref{fig:arrh}. 
Note that the activation energies for the diffusion are divided by a factor of 
two. These activation energies are on the order of $E_{act}=10$~kJ/mol for the 
rotations and on the order of 20~KJ/mol for the diffusion and are only weakly 
dependent on concentration. We see that the in-plane reorientation is stronger 
temperature dependent than the plane flip. At very high temperatures an inversion
of the reorientation order may therefore be expected in line with the moments of 
inertia. This explains the seemingly contradictory finding that the plane-flip is
found to be slower in the simulated range but has the smaller activation energy. 
This is due to the stronger temperature dependence of the in-plane reorientation.

The increasing anisotropy of reorientation with concentration has been reported
for two other systems at higher concentrations. It has been found for water as
solvent of polyvinylacohol~\cite{mplathe98b} and  benzene as solvent of
polystyrene~\cite{mplathe96}.

The diffusion is much stronger temperature dependent than both the rotations. This
can be understood by the fact that for a molecule to diffuse the neighborhood has
to give way in a much stronger fashion, i.e. the whole neighborhood has to 
rearrange whereas for rotation the particle can in principle stay in place.
\section{Conclusion}
Altogether, we showed that the increasing polymer concentration has a
decelerating effect on the dynamics of the cyclohexane solvent molecules.
The polymer chains are obstacles to the cyclohexane motion and make their
reorientation more anisotropic. Additionally we see, in comparison with the
model lacking internal torsions, that even for molecules as small and rigid as
cyclohexane internal torsion potentials have an effect on the overall dynamics.

The simulations presented here were additionally used as starting points for
coarse-graining simulations with investigations of the chain statics and
dynamics~\cite{faller02sa}.
\section*{Acknowledgments}
The author thanks Kevin van Workum for a critical reading of the manuscript.
Financial support by the Emmy-Noether Program of the German Research Foundation
(DFG) is gratefully acknowledged.

\bibliography{standard}
\bibliographystyle{pccp}

\end{document}